\documentclass[aps,pra,amsfonts,twocolumn,amssymb,superscriptaddress,showpacs]{revtex4-1}
\usepackage{graphicx}
\usepackage{dcolumn}
\usepackage{bm,color}
\usepackage{amssymb,amsmath,latexsym}
\allowdisplaybreaks

\begin{document}

\title{Anomalous Dynamic Back-Action in Interferometers}

\author{Sergey~P.~Tarabrin}
\email[Corresponding author:\ ]{Sergey.Tarabrin@aei.mpg.de}
\affiliation{Institut f\"{u}r Gravitationsphysik, Leibniz Universit\"{a}t Hannover and \\ Max-Planck-Institut f\"{u}r Gravitationsphysik (Albert-Einstein-Institut), Callinstra\ss{}e 38, 30167 Hannover, Germany}
\affiliation{Institut f\"ur Theoretische Physik, Leibniz Universit\"at Hannover, Appelstra\ss{}e 2, 30167 Hannover, Germany}

\author{Henning~Kaufer}
\affiliation{Institut f\"{u}r Gravitationsphysik, Leibniz Universit\"{a}t Hannover and \\ Max-Planck-Institut f\"{u}r Gravitationsphysik (Albert-Einstein-Institut), Callinstra\ss{}e 38, 30167 Hannover, Germany}

\author{Farid~Ya.~Khalili}
\affiliation{Moscow State University, Department of Physics, Moscow 119992, Russia}

\author{Roman~Schnabel}
\affiliation{Institut f\"{u}r Gravitationsphysik, Leibniz Universit\"{a}t Hannover and \\ Max-Planck-Institut f\"{u}r Gravitationsphysik (Albert-Einstein-Institut), Callinstra\ss{}e 38, 30167 Hannover, Germany}

\author{Klemens~Hammerer}
\affiliation{Institut f\"{u}r Gravitationsphysik, Leibniz Universit\"{a}t Hannover and \\ Max-Planck-Institut f\"{u}r Gravitationsphysik (Albert-Einstein-Institut), Callinstra\ss{}e 38, 30167 Hannover, Germany}
\affiliation{Institut f\"ur Theoretische Physik, Leibniz Universit\"at Hannover, Appelstra\ss{}e 2, 30167 Hannover, Germany}

\date{\today}

\begin{abstract}
We analyze the dynamic optomechanical back-action in signal-recycled Michelson and Michelson-Sagnac interferometers that are operated off dark port. We show that in this case --- and in contrast to the well-studied canonical form of dynamic back-action on dark port --- optical damping in a Michelson-Sagnac interferometer acquires a non-zero value on cavity resonance, and additional stability/instability regions on either side of the resonance, revealing new regimes of cooling/heating of micromechanical oscillators. In a free-mass Michelson interferometer for a certain region of parameters we predict a stable single-carrier optical spring (positive spring \textit{and} positive damping), which can be utilized for the reduction of quantum noise in future-generation gravitational-wave detectors.
\end{abstract}

\pacs{42.50.Wk, 07.10.Cm, 07.60.Ly, 04.80.Nn}
\maketitle

\section{Introduction}
\begin{figure}
\begin{center}
\includegraphics[scale=0.45]{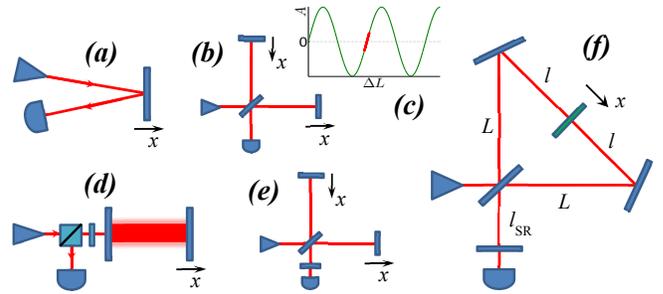}
\caption{Optomechanical setups. (\textit{a}) A simplest setup for measuring the position of the test mass (mirror) via reflected laser beam. (\textit{b}) A Michelson interferometer with end-mirrors performing antisymmetric motion. (\textit{c}) Field amplitude at the output port of a Michelson interferometer as a function of arm-length difference. Points of zero amplitude correspond to dark fringes in the interference pattern. In a small vicinity near dark fringe (marked with thick red) linear signal transduction is possible. (\textit{d}) Resonant amplification of the optical field in a Fabry-Perot cavity. (\textit{e}) A signal-recycled Michelson interferometer. (\textit{f}) A signal-recycled Michelson-Sagnac interferometer with an oscillating semitransparent membrane.}
\label{pic_OM_setups}
\end{center}
\end{figure}

It is a fundamental result of quantum measurement theory that in any optomechanical system, where light serves as a quantum readout agent interacting with a mechanical probe (test mass) via radiation pressure, the probe is subject to measurement \textit{back-action} \cite{Brag_Khalili_1992, Clerk_etal_2010, Danil_Khalili_2012,Aspel_Kipp_Marq_2013,Chen_2013}. Already in the simplest setup comprised of a mirror in ``free-space'', see Fig.\,\ref{pic_OM_setups}\textit{a}, quantum fluctuations of the electromagnetic field exert \textit{back-action noise} on the probe \cite{Caves_1981} which causes the standard quantum limit (SQL) of measurement precision \cite{Brag_1968,Brag_Voron_1975,Brag_Voron_Khalili_1978}. This can become relevant in setups where the optical field is resonantly enhanced, as in a Fabry-Perot (FP) cavity, see Fig.\,\ref{pic_OM_setups}\textit{d}. In this way back-action noise has been observed very recently both  directly \cite{Murch_etal_2008,Purdy2012a} and indirectly via ponderomotive squeezing \cite{Brooks_etal_2012,Safavi-Naeini_etal_2013}. In addition, resonant field enhancement causes \textit{dynamic back-action}: modulation of the intracavity field by the motion of the probe produces a ponderomotive force, which alters the dynamical properties of the probe, as was first recognized in \cite{Brag_Manukin_1967,Brag_Manukin_1970}. Dynamic back-action comprises a shift of the probe's (i) intrinsic damping rate (\textit{optical damping}), and (ii) mechanical frequency (\textit{optical spring}). Both effects have been studied and observed in FP cavities (or equivalent systems) in a regime of \textit{dispersive coupling} (motion of the probe modulates cavity resonance frequency), and utilized for, respectively, optical back-action cooling \cite{Wilson_etal_2007,Marq_etal_2007} of micromechanical oscillators  \cite{Cohadon1999,Metzger2004,Arcizet2006,Gigan2006,Regal2008,Thompson2008,Groeblacher2009,Rocheleau2010,Riviere2011,Teufel2011,Chan2011}, and optical trapping of a gram scale mirror \cite{Corbitt_etal_2006,Corbitt_etal_2007}.

In this paper we address dynamic back-action in standard two-path interferometers, as e.g.~Michelson or Sagnac interferometers, cf.~Fig.\,\ref{pic_OM_setups}. High-precision measurements commonly employ such interferometric topologies because a path differential measurement in a balanced interferometer significantly suppresses path common noise, such as laser noise. The prime example is the Michelson topology of gravitational wave detectors (GWDs). State-of-the-art balanced interferometers utilize resonant field enhancement techniques, similar to FP cavities \cite{Danil_Khalili_2012,McClelland_etal_2011}. For instance, an additional mirror positioned in the interferometer's input and/or output port, referred to as power-recycling (PR) and signal-recycling (SR) mirror respectively, provides amplification of the laser power and/or the signal field \cite{Mee88}. Just as in a FP cavity, the resonant field amplification in recycled interferometers implies dynamic back-action on the test masses. So far, the associated optomechanical effects have been considered exclusively in interferometers perfectly tuned to a dark fringe (vanishing mean power) in the output port (hence, dark port). The operation on \textit{or close to} dark port is in fact the generic working point as it provides the most sensitive signal transduction, see Fig.\,\ref{pic_OM_setups}\textit{c}, and on top of this ensures that laser noise is rejected from the interferometer's output. On dark port the \textit{scaling law} \cite{Buon_Chen_2003} provides a general framework for understanding optomechanics of interferometers: it states that the dynamical and noise properties of any interferometer with high-finesse signal mode and operated on dark port, are equivalent to the ones derived from the dispersive optomechanics of a FP cavity with corresponding effective linewidth, detuning and circulating optical power. Here we show that the dynamic back-action in any SR interferometer operated \textit{off dark port} exhibits rather surprising, anomalous features as compared to the canonical one in a FP cavity as derived within the scope of the scaling law.

We show this for the concrete setups of (i) a Michelson-Sagnac interferometer (MSI) with a semitransparent membrane \cite{Kazuhiro,Friedrich_etal_2011}, see Fig.\,\ref{pic_OM_setups}\textit{f}, as relevant to the optomechanical experiments with micro- and nano-mechanical test masses, and (ii) a free-mass Michelson interferometer (MI), as relevant to GWDs, see Fig.\,\ref{pic_OM_setups}\textit{e}. We emphasize that our logic applies immediately to other interferometric topologies. In particular, we found that optical damping in MSI as a function of detuning acquires a non-zero value on cavity resonance and additional cooling/heating regions on either side of the resonance. In a MI optical spring and damping acquire intersecting regions of positive/negative values.  This is of particular interest for free-mass GWDs as the intricate frequency dependence of the optical spring has been studied (within the scope of scaling law) as a means for increasing the sensitivity via tailoring the dynamics of the test masses \cite{Brag_Gorod_Khalili_1997,Khalili_2001,Rakh_phd_thesis,Buon_Chen_2002,Lazeb_Vyat_2005,Khalili_Lazeb_Vyat_2006}.

From the viewpoint of cavity optomechanics the anomalous dynamic back-action can be understood as being caused by the emergence of \textit{dissipative coupling} (motion of the test mass modulates the cavity linewidth) \cite{Elste_Girv_Clerk_2009,Li2009,Huang2010a,Huang2010,Xuer_Schab_Hammer_2011,Weiss_Burder_Nunnen_2012} and its interplay with the usual dispersive coupling. Recently some of the authors showed that a specially tuned MSI can indeed exhibit a pure dissipative coupling \cite{Xuer_Schab_Hammer_2011}. Here we show that the anomalies in dynamic back-action have to be expected generically for any SR interferometer operated off dark port, even when no simple description in terms of dispersive and dissipative coupling is possible \footnote{The notions of dispersive and dissipative coupling, defined for instance in \cite{Elste_Girv_Clerk_2009}, being unambiguous for a single-mode optomechanical cavity, can become ambiguous for a multi-mode system. In particular, in a two-arm interferometer, being a two-mode system, the type of coupling of the normal mode(s) to the test masses can be different from the one of the partial mode(s)}.

\section{Canonical and anomalous dynamic back-action}\label{sec_canon_anomal_BA}
We derive the dynamic back-action in a SR MSI (which covers MI as a special case) using a transfer matrix approach to the propagation of fields in frequency domain \cite{Danil_Khalili_2012}. Details of calculations are presented in Appendix \ref{sec_propag_fields}. First we characterize the non-recycled MSI as an effective mirror whose reflectance $\rho$ and transmittance $\tau$ depend on the position of the membrane $x=\delta l/2$, with $\delta l$ being the imbalance of the interferometer arms \cite{Xuer_Schab_Hammer_2011}:
\begin{align*}
    \rho&=R_\textrm{m}\left(T_\textrm{BS}^2e^{ik_0\delta l} - R_{\textrm{BS}}^2e^{-ik_0\delta l}\right)-2T_\textrm{m}R_\textrm{BS}T_\textrm{BS},\\
    \tau&=R_\textrm{m}R_\textrm{BS}T_\textrm{BS}\left(e^{ik_0\delta l} + e^{-ik_0\delta l}\right)+T_\textrm{m}(T_\textrm{BS}^2-R_\textrm{BS}^2).
\end{align*}
Here $k_0=\omega_0/c=2\pi/\lambda_0$ is the laser carrier wavenumber, $R_\mathrm{BS}>0$ and $T_\mathrm{BS}>0$ are the amplitude reflectivity and transmissivity of the beamsplitter, and $R_\mathrm{m}>0$ and $T_\mathrm{m}>0$ are the amplitude reflectivity and transmissivity of the membrane. Reflectance $\rho$ describes reflection of the input vacuum field from detector port, while $-\rho^*$ describes reflection of the input laser field. The dark port condition for the interferometer is achieved when $\tau=0$, which is satisfied for $\delta l_\mathrm{DP}=n\lambda_0/4$ and odd $n$ in the case of a balanced beamsplitter, $R_\mathrm{BS}=T_\mathrm{BS}=1/\sqrt{2}$.

Insertion of the SR mirror (SRM) of amplitude reflectivity $R_\mathrm{SR}>0$ and transmissivity $T_\mathrm{SR}>0$ effectively transforms an interferometer into a FP cavity whose second mirror is defined by the MSI. Inverse resonance factor of this effective cavity,
\begin{equation*}
    \mathcal{D}=1-R_\mathrm{SR}|\rho|e^{2ik_0\mathcal{L}+i\arg\rho}=1-R_\mathrm{SR}|\rho|e^{2i\Delta\mathcal{L}/c},
\end{equation*}
where $\mathcal{L}=L+l+l_\mathrm{SR}$ is the effective cavity length (total length between the SRM and the membrane, see Fig.\,\ref{pic_OM_setups}\textit{f}), describes modulation of cavity resonance frequency \textit{and} linewidth by the motion of the membrane, featuring dispersive and dissipative couplings respectively \cite{Xuer_Schab_Hammer_2011}. In particular, total detuning $\Delta$ of laser carrier frequency $\omega_0$ from cavity resonance reads
\begin{align*}
    \Delta&=\delta_\mathrm{SR}+\delta_\mathrm{m},\quad\delta_\mathrm{SR}=\omega_0-\omega_\mathrm{cav},\quad\delta_\mathrm{m}=\frac{c}{2\mathcal{L}}\,\delta\phi,\nonumber\\
    \omega_\mathrm{cav}&=\frac{c}{\mathcal{L}}\pi N-\frac{c}{2\mathcal{L}}\,\arg\rho_{\mathrm{DP}},\quad\delta\phi=\arg\rho-\arg\rho_{\mathrm{DP}}.
\end{align*}
Here $N$ is a fixed integer, $\rho_\mathrm{DP}=\rho(\delta l_\mathrm{DP})$, $\delta_\mathrm{SR}$ is the detuning from cavity resonance $\omega_\mathrm{cav}$ at dark port, defined by the position of the SRM, and $\delta_\mathrm{m}$ is the detuning due to offset from dark port, defined by the position of the membrane, hence, describing the dispersive coupling in the effective cavity. Similarly, total half-linewidth $\gamma$ of the latter in the narrow-band approximation ($T_\mathrm{SR}\ll1,\ \tau\ll1$),
\begin{equation*}
    \gamma=\frac{1-R_\mathrm{SR}|\rho|}{2\mathcal{L}/c}=\gamma_\mathrm{SR}+\gamma_\mathrm{m},\
    \gamma_\mathrm{SR}=\frac{cT_\mathrm{SR}^2}{4\mathcal{L}},\
    \gamma_\mathrm{\mathrm{m}}=\frac{c\tau^2}{4\mathcal{L}},
\end{equation*}
accounts for the finite transmittances of the SRM via $\gamma_\mathrm{SR}$, and of the MSI operated off dark port via $\gamma_\mathrm{m}$, the latter thus describing dissipative coupling in the effective cavity.

We then compute the fields incident on the membrane and derive the corresponding radiation pressure force, which consists of (i) the radiation pressure noise, that is stochastic back-action due to vacuum fluctuations entering from laser and detector ports, and (ii) the ponderomotive force due to motion of the membrane $x(\Omega)$. The unsymmetrized spectral density of back-action noise exhibits a mixture of Lorentz- and Fano-like resonances, the latter owned to the interference between input and intracavity laser fields, see Refs.\,\cite{Elste_Girv_Clerk_2009,Xuer_Schab_Hammer_2011} and Appendix \ref{sec_stochastic_BA}. The ponderomotive force $F_x(\Omega)=-\mathcal{K}(\Omega)x(\Omega)$, calculated in Appendix \ref{sec_dynamic_BA}, creates a dynamic back-action $\mathcal{K}(\Omega)$, comprised of the optical spring $K(\Omega)=\Re[\mathcal{K}(\Omega)]$ and optical damping $\Gamma(\Omega)=-\frac{1}{2}\Im[\mathcal{K}(\Omega)]/\Omega$, such that the corresponding shifts of the square of the mechanical frequency and mechanical damping rate are $K/m$ and $\Gamma/m$, with $m$ being the membrane's (effective) mass.

\begin{figure}
\begin{center}
\includegraphics[scale=0.45]{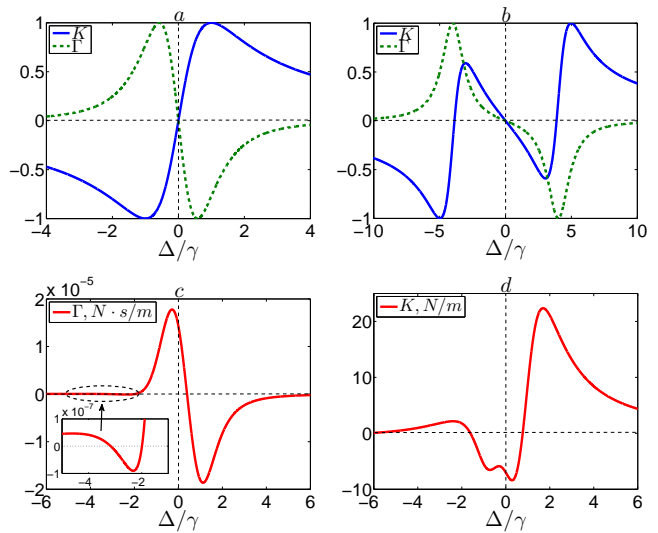}
\caption{
(\textit{a}) Canonical $K$ and $\Gamma$ defined in Eq.\,(\ref{eq_dynamic_BA_gen_approx}) for $\xi=0$, normalized to their respective maximum values, for $\Omega<\gamma$. (\textit{b}) Same for $\Omega>\gamma$. (\textit{c}) and (\textit{d}) respectively show optical damping and spring in the signal-recycled MSI detuned from dark port at $\delta l_\mathrm{DP}=3\lambda_0/4$ by $\xi\approx0.01\lambda_0$, with $P_\mathrm{in}=200$ mW, $\lambda_0=1064$ nm, $\mathcal{L}=8.7$ cm, $R_\mathrm{m}^2=0.17$, $T_\mathrm{SR}^2=3\times10^{-4}$ and $\gamma\approx\Omega=2\pi\cdot133$ kHz.
}
\label{pic_BA}
\end{center}
\end{figure}

We present the formula for $\mathcal{K}(\Omega)$, which is rather complicated in the general case, using the following simplifying assumptions: (i) balanced beamsplitter, (ii) single optical mode with narrow linewidth, $\gamma\ll c/\mathcal{L}$, and (iii) small displacements of the membrane from the position corresponding to dark fringe, $\langle x\rangle=\xi/2\ll\lambda_0$, with $\xi=\delta l-\delta l_\mathrm{DP}$. Under these conditions, $\gamma_\mathrm{m}=cR_\mathrm{m}^2(k_0\xi)^2/4\mathcal{L}$, $\delta_\mathrm{m}=\pm cR_\mathrm{m}T_\mathrm{m}(k_0\xi)^2/4\mathcal{L}$~\footnote{The sign of $\delta_\mathrm{m}$ alternates in the sequence of dark ports, starting with '$+$' for $n=1$. Note that both shifts in linewidth ($\gamma_\mathrm{m}$) and detuning ($\delta_\mathrm{m}$) due to offset from dark fringe are \textit{quadratic} in displacement.} and the dynamic back-action reads
\begin{multline}
    \mathcal{K}(\Omega)=\frac{4\omega_0R_\mathrm{m}^2P_\mathrm{in}}{c\mathcal{L}}\,\frac{1}{\Delta^2+(\gamma-i\Omega)^2}\\
    \times\biggl\{\frac{\delta_\mathrm{SR}[\gamma^2+\Delta^2-4(\gamma\gamma_\mathrm{m}+\Delta\delta_\mathrm{m})]}{\gamma^2+\Delta^2}\\
    +\frac{2i(\gamma_\mathrm{SR}\delta_\mathrm{m}+\gamma_\mathrm{m}\delta_\mathrm{SR})\Omega+\delta_\mathrm{m}\Omega^2}{\gamma^2+\Delta^2}\biggr\},
    \label{eq_dynamic_BA_gen_approx}
\end{multline}
where $P_\mathrm{in}$ is the input laser power. For $\xi=0$ (on dark port) Eq.\,(\ref{eq_dynamic_BA_gen_approx}) reduces to the canonical spring and damping of a simple FP cavity with pure dispersive coupling \cite{Khalili_2001,Buon_Chen_2003}. These canonical $K$ and $\Gamma$ posses the following characteristic features: (i) both are antisymmetric with respect to $\Delta$ and vanish at $\Delta=0$, (ii) $\Gamma$ as a function of $\Delta$ crosses zero only once, being positive for $\Delta<0$ and negative for $\Delta>0$ --- these regions are usually labeled as stable (cooling) and unstable (heating), (iii) $K$ as a function of $\Delta$ crosses zero once if $\gamma\geq\Omega$ (case of free-mass GWD interferometers) and (iv) three times otherwise (case of micromechanical oscillators in the resolved sideband limit). These properties are illustrated in Fig.\,\ref{pic_BA}\textit{a,b}.

Off dark port for $\xi\neq0$, features (i) --- (iii) break. In this sense we refer to the dynamic back-action in a MSI operated off dark fringe as anomalous. In particular, both $K$ and $\Gamma$ become highly asymmetric and acquire non-zero values at $\Delta=0$, see Fig.\,\ref{pic_BA}\textit{c},\textit{d}, such that for a certain region of parameters $\Gamma|_{\Delta=0}>0$ --- this is \textit{cooling on cavity resonance}. Also optical damping can cross zero several times, acquiring additional regions of stability/instability (see inset in Fig.\,\ref{pic_BA}\textit{c}), thus allowing \textit{another regime of cooling/heating}. Non-zero $K$ at $\Delta=0$ implies a shift of the mechanical frequency on resonance, although for micromechanical oscillators it is mostly negligible compared to their intrinsic mechanical frequencies. The influence of the optical losses can be estimated by corresponding increase of $T_\mathrm{SR}^2$.

We note that at fixed mechanical frequency $\omega_\mathrm{m}$ optical damping is proportional to the antisymmetric part of the unsymmetrized spectral density $S_F$ of back-action noise, $\Gamma\sim S_F(\omega_\mathrm{m})-S_F(-\omega_\mathrm{m})$, see \cite{Clerk_etal_2010}. Since in cavity optomechanics the transformation of the Lorentzian profile of $S_F$ into the mixture of Lorentz and Fano ones is governed by the interplay between dispersive and dissipative couplings \cite{Elste_Girv_Clerk_2009,Xuer_Schab_Hammer_2011,Weiss_Burder_Nunnen_2012}, one can argue that the same mechanism leads to the transformation of canonical dynamic back-action into the anomalous one. Dynamic back-action corresponding to the pure dissipative coupling in a single-mode cavity optomechanics was considered in \cite{Weiss_Burder_Nunnen_2012}, where it was shown that although both optical spring and damping remain antisymmetric with respect to $\Delta$ (and vanish at $\Delta=0$), damping acquires additional regions of stability/instability on either side of the cavity resonance.

An extreme case of a 100\% reflective membrane in MSI corresponds to a pure MI, i.e.~reproduces basic topology of the GWDs. The coordinate $x$ of the mechanical degree of freedom refers then to the differential motion of the end-mirrors in the arms of the MI, see Fig.\,\ref{pic_OM_setups}\textit{e}. For a GWD being a free-mass interferometer, the effect of the optical spring is not negligible, since it transforms (almost) free test masses into mechanical oscillators with resonance frequencies lying in the GW observation band, where typically $\Omega<\gamma$. Thus, if a detuned interferometer is operated at dark fringe, Eq.\,(\ref{eq_dynamic_BA_gen_approx}) for $\xi=0$ implies either $K>0$, $\Gamma<0$ for $\Delta>0$, or $K<0$, $\Gamma>0$ for $\Delta<0$. This means that for a single laser drive a set of canonical $K$ and $\Gamma$ is unstable in both cases.

\begin{figure}
\begin{center}
\includegraphics[scale=0.45]{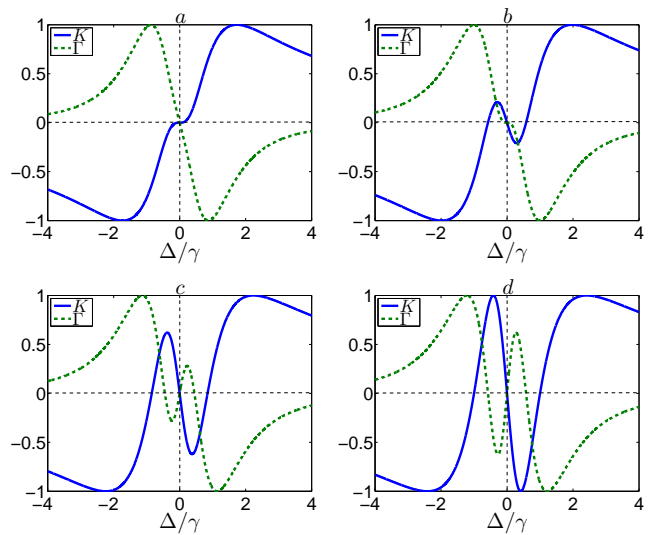}
\caption{Anomalous $K$ and $\Gamma$ in a Michelson interferometer, defined by Eqs.\,(\ref{eq_optical spring_anom},\,\ref{eq_optical damping_anom}), and normalized to their respective maximum values. (\textit{a}) $K$ acquires three zeroes starting from $\gamma_\mathrm{m}=\gamma/4$, or equivalently, $\gamma_\mathrm{m}=\gamma_\mathrm{SR}/3$. (\textit{b}) $\Gamma$ acquires three zeroes starting from $\gamma_\mathrm{m}=\gamma/3$ or $\gamma_\mathrm{m}=\gamma_\mathrm{SR}/2$. (\textit{c}) $\gamma_\mathrm{m}=7\gamma/3$ or $\gamma_\mathrm{m}=3\gamma_\mathrm{SR}/4$, and (\textit{d}) $\gamma_\mathrm{m}=\gamma/2$ or $\gamma_\mathrm{m}=\gamma_\mathrm{SR}$ demonstrate the existence of stable optical spring ($K>0$ and $\Gamma>0$) in a certain range of detunings for larger offsets from dark fringe.}
\label{pic_BA_Michelson}
\end{center}
\end{figure}

However, a MI also exhibits anomalous dynamic back-action if operated off dark port, violating features (ii) and (iii) of the canonical one: in the limit of a quasi free mass, $\Omega\rightarrow0$, which is of particular interest for GWDs, Eq.\,(\ref{eq_dynamic_BA_gen_approx}) reduces to
\begin{subequations}
\begin{align}
    K&=\frac{4\omega_0P_\mathrm{in}}{c\mathcal{L}}\,\frac{\Delta}{\gamma^2+\Delta^2}\,\left[1-\frac{4\gamma\gamma_\mathrm{m}}{\gamma^2+\Delta^2}\right],\label{eq_optical spring_anom}\\
    \Gamma&=-\frac{4\omega_0P_\mathrm{in}}{c\mathcal{L}}\,\frac{\gamma\Delta}{(\gamma^2+\Delta^2)^2}\,
    \left[1-\frac{\gamma_\mathrm{m}}{\gamma}\,\frac{3\gamma^2-\Delta^2}{\gamma^2+\Delta^2}\right].\label{eq_optical damping_anom}
\end{align}
\end{subequations}
Both $K$ and $\Gamma$ vanish on resonance, and one can check using Eq.\,(\ref{eq_dynamic_BA_gen_approx}) that this feature holds for any $\Omega$. Terms in square brackets in Eqs.\,(\ref{eq_optical spring_anom},\,\ref{eq_optical damping_anom}) represent the deviations from canonical formulas. According to these terms, the optical spring can have three zeroes at $\Delta=0$ and $\Delta=\pm\sqrt{\gamma(4\gamma_\mathrm{m}-\gamma)}$, if $\gamma_\mathrm{m}>\gamma/4$. Similarly, the optical damping can also cross zero three times at $\Delta=0$ and $\Delta=\pm\gamma\sqrt{(3\gamma_\mathrm{m}-\gamma)/(\gamma+\gamma_\mathrm{m})}$, if $\gamma_\mathrm{m}>\gamma/3$. This sequence of transformations of the canonical $K$ and $\Gamma$, shown in Fig\,\ref{pic_BA}\textit{a}, into the anomalous ones for the increasing value of $\gamma_\mathrm{m}$ is illustrated in Fig.\,\ref{pic_BA_Michelson}. Since the 2nd-generation GWDs will be utilizing the single-photodiode homodyne readout (DC-readout), when the offset from dark port is created on purpose to transmit a small portion of mean power to the detection port, anomalous optical spring/damping may have an impact on the control of detectors in detuned regime ($\Delta\neq0$). Additionally, in realistic dual-recycled interferometers (SR \textit{and} PR) anomalous optical spring should be expected at even smaller $\xi$, since anomalies rise at $k_0\xi\sim\sqrt{T_\mathrm{SR}^2T_\mathrm{PR}^2}$, where $T_\mathrm{PR}^2$ stands for the power reflectivity of the PR mirror, compared to $k_0\xi\sim\sqrt{T_\mathrm{SR}^2}$ for pure SR topologies.

Fig.\,\ref{pic_BA_Michelson}\textit{c,d} shows that for a large enough offset from dark fringe several intersecting regions of positive/negative $K$ and $\Gamma$ appear, such that for a certain range of negative detunings both $K$ and $\Gamma$ are positive, indicating a possible stable state. Accurate analysis of stability in terms of Routh-Hurwitz criteria indeed reveals that there exists a region of parameters where the set of $K$ and $\Gamma$ makes a stable optical spring which can be utilized for the increase of the quantum-noise-limited sensitivity of GWD topologies, such as traditional detuned-SR topology and promising intracavity topologies of optical bars \cite{Khalili_2003} and optical levers \cite{Khalili_2003,Danil_Khalili_2006}.

\section{Conclusions}
We have shown that the dynamic back-action in interferometers operated off dark port exhibits anomalous features as compared to the canonical one in dark-port-operated interferometers, and should be expected as a generic feature in optomechanical systems which exhibit a mixture of dispersive and dissipative couplings. In particular, in a SR MSI with a translucent micromechanical membrane, cooling/heating of the latter becomes possible on resonance. Additionally, as a generic feature of dissipatively coupled systems, cooling of the membrane to near quantum ground state outside the resolved sideband regime can be in principle achieved. We have demonstrated that the scaling law --- being the cornerstone of characterization of the optomechanics of GWD topologies --- is invalidated for interferometers operated off dark port. In particular, for a large enough offset from dark port in a MI, a stable optical spring, used for the reduction of quantum noise, becomes possible with a single laser carrier. The latter condition is important, because the canonical optical spring (on dark port) can only be stabilized with either additional feedback/control loops or second laser carrier \cite{Rehb_etal_2008,Rakhub_Hild_Vyat_2011,Rakhub_Vyat_2012}, which is undesirable in the experiment. Thus, the offset from dark fringe makes a useful additional degree of freedom in shaping of the optical spring, and, in principle, may help in converting the latter into the so-called \textit{optical inertia} \cite{Khalili_etal_2011} which allows broadband reduction of quantum noise. However, since a large offset from dark fringe will couple mean power and technical laser noise into the detector port, certain changes in the GWD topology will be required to make use of the stable single-carrier optical spring, such e.g.~switching to intrcavity topologies \cite{Khalili_2003,Danil_Khalili_2006}. The anomalous aspects of back-action introduced in this paper, which will be important to estimate the consequences for next-generation GWDs, can be examined and studied on the basis of a micromechanical system. Therefore, our findings represent an example of the fruitful interplay between cavity optomechanics of micromechanical oscillators and the optomechanics of ``macro-mechanical'' oscillators as relevant to the high-precision interferometers employed for GWD.

\acknowledgments
We would like to thank Karsten Danzmann, Harald L\"{u}ck, Yanbei Chen and Stefan Hild for fruitfull discussions. This work was funded by the Centre for Quantum Engineering and Space-Time Research (QUEST) at the Leibniz University Hannover, and by the European Commission through iQUOEMS. F.~K.~was supported by LIGO NSF grant PHY-0967049 and Russian Foundation for Basic Research grant No.11-02-00383-a.

\appendix
\section{Propagation of fields}\label{sec_propag_fields}

\begin{figure}
\begin{center}
\includegraphics[scale=0.5]{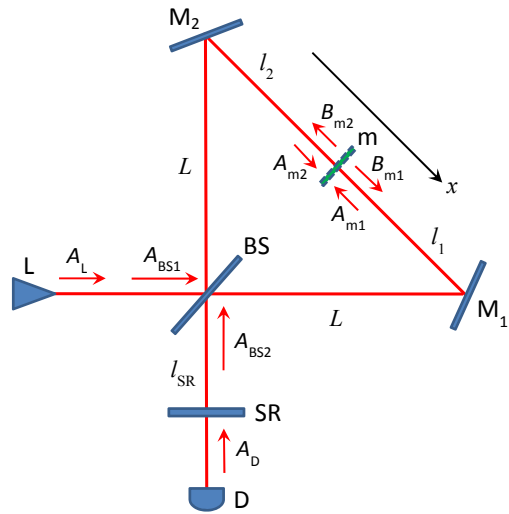}
\caption{Fields in a Michelson-Sagnac interferometer.}
\label{pic_MSR_interferometer}
\end{center}
\end{figure}

Consider a Michelson-Sagnac interferometer (MSI) as shown in Fig.\,\ref{pic_MSR_interferometer} with a central beamsplitter BS having amplitude reflectivity $R_\mathrm{BS}=\sqrt{(1-\delta_{\mathrm{BS}})/2}$ and transmissivity $T_{\mathrm{BS}}=\sqrt{(1+\delta_\mathrm{BS})/2}$, where $0\leq\delta_\mathrm{BS}\leq1$, two steering mirrors $\textrm{M}_1$ and $\textrm{M}_2$ both having 100\% reflectivity, a semitransparent membrane m with amplitude reflectivity $R_\mathrm{m}>0$ and transmissivity $T_{\mathrm{m}}>0$, and a signal-recycling mirror SR with amplitude reflectivity $R_\mathrm{SR}>0$ and transmissivity $T_{\mathrm{SR}}>0$. The interferometer is driven by a laser L through laser port. Photons emanating through the other, detector port impinge on a detector D (homodyne or heterodyne). We denote the distance between SR mirror and BS as $l_\mathrm{SR}$, arm length as $L$ and the distances between folding mirrors $\textrm{M}_1$ and $\textrm{M}_2$ and membrane as $l_1 = l-\delta l/2$ and $l_2 = l+\delta l/2$, respectively. This means that $l_1+l_2=2l$, $l_2-l_1=\delta l$ and the mean position of the membrane on the $x$-axis is $\langle x \rangle=\delta l/2$. The total length of the SR-m path is $\mathcal{L}=L+l+l_\mathrm{SR}$.

In any spatial location inside the interferometer we decompose the electric field of the coherent, plane and linearly polarized electromagnetic wave into the sum of a steady-state (mean) field with amplitude $A_0$ and carrier frequency $\omega_0$ (wavenumber $k_0=\omega_0/c$ and wavelength $\lambda_0=2\pi/k_0$), and slowly-varying (on the scale of $1/\omega_0$) perturbation field with amplitude $a(t)$ describing vacuum noises and the contribution from the motion of the membrane,
\begin{align*}
    A(t)&=\sqrt{\frac{2\pi\hbar\omega_0}{\mathcal{A}c}}\left[A_0e^{-i\omega_0t} + a(t)e^{-i\omega_0t}\right] + \textrm{{h.c.}},\\
    a(t)&=\int_{-\infty}^{+\infty}a(\omega_0+\Omega)e^{-i\Omega t}\,\frac{d\Omega}{2\pi}.
\end{align*}
Here $\mathcal{A}$ is the area of laser beam's cross-section and $c$ is the speed of light. Unless mentioned explicitly, we will deal with fields in the frequency domain only and omit frequency arguments for briefness.

The laser L emits a drive-wave $A_\mathrm{L}$ with mean amplitude $A_{\mathrm{L}0}$ and optical fluctuations $a_\mathrm{L}$. For simplicity we assume that there are no technical fluctuations so that the laser is shot-noise limited, $[a_\mathrm{L}(\omega_0+\Omega),a_\mathrm{L}^\dag(\omega_0+\Omega')]=2\pi\delta(\Omega-\Omega')$. The vacuum field $A_\mathrm{D}$ entering through the SR mirror (SRM) from detector port has zero mean amplitude but non-zero vacuum noise $a_\mathrm{D}$, uncorrelated with vacuum noise from the laser port and obeying the similar commutation relation $[a_\mathrm{D}(\omega_0+\Omega),a_\mathrm{D}^\dag(\omega_0+\Omega')]=2\pi\delta(\Omega-\Omega')$. We unite these into vector-column of input fields $\textit{\textbf{A}}_\mathrm{in}=(A_\mathrm{L},A_\mathrm{D})$, so that the vector of mean input fields is $\textit{\textbf{A}}_{\mathrm{in}0}=(A_{\mathrm{L}0},0)$ and the vector of perturbation fields is $\textit{\textbf{a}}_\mathrm{in}=(a_\mathrm{L},a_\mathrm{D})$. Due to linearity of the system input fields can be propagated throughout the interferometer as independent Fourier components.

Consider first the case without SRM and with a fixed membrane. The latter condition allows us to treat mean and perturbation fields on equal footing. Input fields (in this case coinciding with the fields incident on the beamsplitter) linearly transform into the output fields: $\textit{\textbf{A}}_{\mathrm{out}}=\mathbb{M}_\textrm{BS}^T\mathbb{P}_L\mathbb{P}_l\mathbb{M}_\textrm{m}\mathbb{P}_l\mathbb{P}_L\mathbb{M}_\textrm{BS}\textit{\textbf{A}}_{\mathrm{in}}
\equiv\mathbb{M}_\textrm{MS}\textit{\textbf{A}}_{\mathrm{in}}$. Here
\begin{equation}
    \mathbb{M}_\mathrm{BS}=
    \begin{pmatrix}
      T_\textrm{BS} & -R_\textrm{BS} \\
      R_\textrm{BS} & T_\textrm{BS} \\
    \end{pmatrix},\quad
    \mathbb{M}_\textrm{m}=
    \begin{pmatrix}
      -R_\textrm{m} & T_\textrm{m} \\
      T_\textrm{m} & R_\textrm{m} \\
    \end{pmatrix},
    \label{eq_bs_membr_transfer_matrices}
\end{equation}
are the transformation matrices of beamsplitter and membrane, both chosen in real form (this is always possible due to Stokes relations), and
\begin{equation*}
    \mathbb{P}_L=
    \begin{pmatrix}
      e^{ikL} & 0 \\
      0 & e^{ikL} \\
    \end{pmatrix},\quad
    \mathbb{P}_l=
    \begin{pmatrix}
      e^{ikl_1} & 0 \\
      0 & e^{ikl_2} \\
    \end{pmatrix},
\end{equation*}
are the propagation matrices comprised of the phase shifts along the horizontal/vertical arms (of length $L$) and diagonal half-arms (of lengths $l_{1,2}$). For mean fields one should apply the substitution $k=k_0$ and for perturbation fields $k=k_0+K=k_0+\Omega/c$. The matrix $\mathbb{M}_\textrm{MS}$ thus represents the transformation matrix of a non-recycled MSI
\begin{equation*}
    \mathbb{M}_\textrm{MS}=e^{2ik(L+l)}
    \begin{pmatrix}
      \rho_1 & \tau \\
      \tau & \rho_2 \\
    \end{pmatrix},
\end{equation*}
with
\begin{align*}
    \rho_1&=R_\textrm{m}\left(R_\textrm{BS}^2e^{ik\delta l} - T_{\textrm{BS}}^2e^{-ik\delta l}\right)+
    2T_\textrm{m}R_\textrm{BS}T_\textrm{BS},\\
    \rho_2&=R_\textrm{m}\left(T_\textrm{BS}^2e^{ik\delta l} - R_{\textrm{BS}}^2e^{-ik\delta l}\right)-
    2T_\textrm{m}R_\textrm{BS}T_\textrm{BS},\\
    \tau&=R_\textrm{m}R_\textrm{BS}T_\textrm{BS}\left(e^{ik\delta l} + e^{-ik\delta l}\right)+
    T_\textrm{m}(T_\textrm{BS}^2-R_\textrm{BS}^2).
\end{align*}
In Sec.\,\ref{sec_canon_anomal_BA} we denoted $\rho\equiv\rho_2$. Physically $\rho_1$ is the reflectivity of the input laser field back into the laser port, $\rho_2=-\rho_1^*$ is the reflectivity of input vacuum field back into detector port, and $\tau$ is the transmissivity of the laser field into detector port and vacuum field into laser port. One can check that the matrix $\mathbb{M}_\textrm{MS}$ is unitary, thus \textit{non-recycled MSI can be described as an effective mirror with reflectivity and transmissivity depending on membrane position} via $\delta l$, as stated in Sec.\,\ref{sec_canon_anomal_BA} and discussed already in \cite{Xuer_Schab_Hammer_2011}. The dark port (dark fringe) condition for the interferometer is achieved when the cross-transmittance between input-output ports vanishes (in particular, no mean power leaks into the detector port), corresponding to $\tau=0$, or explicitly
\begin{equation}
    \cos k_0\delta l=-\frac{T_\textrm{m}}{R_\textrm{m}}\,\frac{\delta_\textrm{BS}}{\sqrt{1-\delta_\textrm{BS}^2}}.
    \label{eq_dark_port_condition}
\end{equation}
In the case of a symmetric beamsplitter ($\delta_\textrm{BS}=0$) this is satisfied for $\delta l=n\lambda_0/4$ and odd $n$ \footnote{For other choices of transfer matrices (\ref{eq_bs_membr_transfer_matrices}) the dark port condition will correspond to a different $\delta l$ but this is insignificant, since the absolute phase does not matter}.

If the SRM is inserted then the out-going field in the SR port is reflected back, such that the in-going fields incident on the beamsplitter are defined by the equation
\begin{equation}
    \textit{\textbf{A}}_\textrm{BS}=\mathbb{P}_\textrm{R}\mathbb{T}_\textrm{R}\textit{\textbf{A}}_{\textrm{in}}+
    \mathbb{P}_\textrm{R}\mathbb{R}_\textrm{R}\mathbb{P}_\textrm{R}\mathbb{M}_\textrm{MS}\textit{\textbf{A}}_\textrm{BS}.
    \label{eq_bs_fields_immov_membr}
\end{equation}
Here $\textit{\textbf{A}}_{\textrm{BS}}=(A_{\textrm{BS}1},A_{\textrm{BS}2})$ is the vector-column of in-going beamsplitter fields (see Fig.\,\ref{pic_MSR_interferometer}), $\mathbb{R}_\textrm{R}=\mathrm{diag}(0,R_\textrm{SR})$ with zero standing for the absence of power-recycling mirror in laser port, $\mathbb{P}_\textrm{R}=\mathrm{diag}(1,e^{ikl_\textrm{SR}})$ is the propagation matrix in BS-SR path, and $\mathbb{T}_\textrm{R}=\mathrm{diag}(1,T_\textrm{SR})$. Thus the first summand on the RHS of Eq.\,(\ref{eq_bs_fields_immov_membr}) stands for the input fields directly incident on the beamsplitter, while the second summand corresponds to a single round trip along the interferometer with reflection from the SRM. Solution of this equation yields
\begin{equation}
    \textit{\textbf{A}}_\textrm{BS}=(\mathbb{I}-\mathbb{P}_\textrm{R}\mathbb{R}_\textrm{R}\mathbb{P}_\textrm{R}\mathbb{M}_\textrm{MS})^{-1}\mathbb{P}_\textrm{R}\mathbb{T}_\textrm{R}
    \textit{\textbf{A}}_{\textrm{in}},
    \label{eq_bs_ingoing_fields_immov_membr}
\end{equation}
where $\mathbb{I}$ is the $2\times2$ unity matrix. Denote inverse matrix in this solution as $\mathbb{K}_{\textrm{MSR}}$,
\begin{equation*}
    \mathbb{K}_{\textrm{MSR}}=\frac{1}{\mathcal{D}}
    \begin{pmatrix}
      \mathcal{D} & 0 \\
      R_\textrm{SR}\tau e^{2ik\mathcal{L}} & 1 \\
    \end{pmatrix},\quad
    \mathcal{D}=1-R_\mathrm{SR}\rho_2e^{2ik\mathcal{L}}.
\end{equation*}
This tells us that \textit{the MSI with SRM makes an effective Fabry-Perot cavity} with associated resonance factor $1/\mathcal{D}$, as stated in Sec.\,\ref{sec_canon_anomal_BA}. The matrix element $\mathbb{K}_{\textrm{MSR}}^{(2,2)}$ describes a resonant amplification of input vacuum field inside the cavity, while $\mathbb{K}_{\textrm{MSR}}^{(2,1)}$ corresponds to the laser field being partially transmitted into the SR port (hence the proportionality to $\tau$) and also enhanced inside the cavity. In the ideal dark-port regime cross-transmittance is suppressed, all laser field is reflected back into laser port, and only the vacuum field from the detector port resonates inside the cavity.

Note that the effective detuning of the laser carrier from cavity resonance(s) is not solely defined by the corresponding shift in frequency (or cavity length) in contrast to the ordinary Fabry-Perot cavity. Assume that $\arg\rho_2=\phi_{\textrm{DP}}+\delta\phi$, where $\phi_\mathrm{DP}=\arg\rho_2|_{\textrm{dark port}}$ and $\delta\phi$ is the deviation from it due to offset from dark fringe via membrane positioning, and $2k\mathcal{L}+\phi_{\textrm{DP}}=2\pi N+2\delta_k\mathcal{L}$, $N$ is integer, $\delta_k\mathcal{L}\ll1$. Then one can rewrite the inverse resonance factor as
\begin{align*}
    \mathcal{D}&=1-R_\mathrm{SR}|\rho_2|e^{2ik\mathcal{L}+i\arg\rho_2}=1-R_\mathrm{SR}|\rho_2|e^{2i\Delta\mathcal{L}/c},
\end{align*}
such that the full detuning
\begin{equation*}
    \Delta=\delta+\frac{c}{2\mathcal{L}}\,\delta\phi,
\end{equation*}
is the sum of the 'conventional' detuning $\delta=c\delta_k$ of carrier frequency from cavity resonance at dark port, and an additional detuning $c\delta\phi/(2\mathcal{L})$ corresponding to the offset from the latter.

The narrow-band limit is achieved when both SRM and compound 'interferometer' mirror possess high reflectivity, $1-R_\mathrm{SR}\approx T_\mathrm{SR}^2/2\ll1$ and $1-|\rho_2|\approx\tau^2/2\ll1$. The half-linewidth of the cavity is then
\begin{equation*}
    \gamma=\frac{1-R_\mathrm{SR}|\rho_2|}{2\mathcal{L}/c}\approx\frac{cT_\mathrm{SR}^2}{4\mathcal{L}}+\frac{c\tau^2}{4\mathcal{L}}.
\end{equation*}
Therefore, the total cavity linewidth accounts for finite SRM transmittance and finite transmittance of the interferometer operated off dark port; since $\tau=\tau(\delta l)$, the latter contribution describes modulation of the linewidth by the motion of the membrane, thus implementing dissipative coupling in the effective cavity, as stated in Sec.\,\ref{sec_canon_anomal_BA} and discussed already in \cite{Xuer_Schab_Hammer_2011}.

\section{Stochastic back-action}\label{sec_stochastic_BA}

In order to determine the radiation pressure force acting on the membrane we need to determine the fields on the membrane surfaces. In-going fields on the beamsplitter (\ref{eq_bs_ingoing_fields_immov_membr}) propagate along the arms and transform into the fields incident on the membrane $(A_{\textrm{m}1},A_{\textrm{m}2})=\textit{\textbf{A}}_\textrm{m}=\mathbb{P}_l\mathbb{P}_L\mathbb{M}_\textrm{BS}\textit{\textbf{A}}_\textrm{BS}$ and reflected from it $(B_{\textrm{m}1},B_{\textrm{m}2})=\textit{\textbf{B}}_\textrm{m}=\mathbb{M}_\textrm{m}\textit{\textbf{A}}_\textrm{m}$, see Fig.\,\ref{pic_MSR_interferometer}. In terms of input fields
\begin{subequations}
\begin{align}
    \textit{\textbf{A}}_\textrm{m}&=\mathbb{M}_A\textit{\textbf{A}}_\textrm{in};\
    \mathbb{M}_A=\mathbb{P}_l\mathbb{P}_L\mathbb{M}_\textrm{BS}\mathbb{K}_{\textrm{MSR}}\mathbb{P}_\textrm{R}\mathbb{T}_\textrm{R},\label{eq_membr_fields_inc_immov_membr}\\
    \textit{\textbf{B}}_\textrm{m}&=\mathbb{M}_B\textit{\textbf{A}}_\textrm{in};\
    \mathbb{M}_B=\mathbb{M}_\mathrm{m}\mathbb{P}_l\mathbb{P}_L\mathbb{M}_\textrm{BS}\mathbb{K}_{\textrm{MSR}}\mathbb{P}_\textrm{R}\mathbb{T}_\textrm{R}.
    \label{eq_membr_fields_ref_immov_membr}
\end{align}
\end{subequations}
The components of matrix $\mathbb{M}_A$ are:
\begin{align*}
    \mathbb{M}_A^{(1,1)}={}&\mathcal{D}^{-1}\Bigl[T_\textrm{BS}\left(1-R_\textrm{m}R_\textrm{SR}e^{2ik(\mathcal{L}+\delta l/2)}\right)\nonumber\\
    &+R_\textrm{BS}T_\textrm{m}R_\textrm{SR}e^{2ik\mathcal{L}}\Bigr]e^{ik(L+l-\delta l/2)},\\
    \mathbb{M}_A^{(1,2)}={}&-\mathcal{D}^{-1}T_\textrm{SR}R_\textrm{BS}e^{ik(\mathcal{L}-\delta l/2)},\\
    \mathbb{M}_A^{(2,1)}={}&\mathcal{D}^{-1}\Bigl[R_\textrm{BS}\left(1+R_\textrm{m}R_\textrm{SR}e^{2ik(\mathcal{L}-\delta l/2)}\right)\nonumber\\
    &+T_\textrm{BS}T_\textrm{m}R_\textrm{SR}e^{2ik\mathcal{L}}\Bigr]e^{ik(L+l+\delta l/2)},\\
    \mathbb{M}_A^{(2,2)}={}&\mathcal{D}^{-1}T_\textrm{SR}T_\textrm{BS}e^{ik(\mathcal{L}+\delta l/2)},
\end{align*}
and of matrix $\mathbb{M}_B$:
\begin{align*}
    \mathbb{M}_B^{(1,1)}={}&\mathcal{D}^{-1}\Bigl[
    -T_\textrm{BS}\left(R_\textrm{m}-R_\textrm{SR}e^{2ik(\mathcal{L}+\delta l/2)}\right)\nonumber\\&+T_\textrm{m}R_\textrm{BS}e^{ik\delta l}\Bigr]e^{ik(L+l-\delta l/2)},\\
    \mathbb{M}_B^{(1,2)}={}&\mathcal{D}^{-1}T_\textrm{SR}\left(R_\textrm{BS}R_\textrm{m}+T_\textrm{m}T_\textrm{BS}e^{ik\delta l}\right)e^{ik(\mathcal{L}-\delta l/2)},\\
    \mathbb{M}_B^{(2,1)}={}&\mathcal{D}^{-1}\Bigl[
    R_\textrm{BS}\left(R_\textrm{m}+R_\textrm{SR}e^{2ik(\mathcal{L}-\delta l/2)}\right)\nonumber\\&+T_\textrm{m}T_\textrm{BS}e^{-ik\delta l}\Bigr]e^{ik(L+l+\delta l/2)},\\
    \mathbb{M}_B^{(2,2)}={}&\mathcal{D}^{-1}T_\textrm{SR}\left(T_\textrm{BS}R_\textrm{m}-T_\textrm{m}R_\textrm{BS}e^{-ik\delta l}\right)e^{ik(\mathcal{L}+\delta l/2)}.
\end{align*}
Denote these transfer matrices separately for mean fields as $\mathbb{M}_{A_0}=\mathbb{M}_A|_{k=k_0}$, $\mathbb{M}_{B_0}=\mathbb{M}_B|_{k=k_0}$ and perturbation fields as $\mathbb{M}_{a}(\Omega)=\mathbb{M}_A|_{k=k_0+K}$, $\mathbb{M}_{b}(\Omega)=\mathbb{M}_B|_{k=k_0+K}$.

The radiation pressure force exerted on the membrane is then given by
\begin{equation}
    F(t)=-\frac{\mathcal{A}}{4\pi}\left\langle A_{\textrm{m}1}^2(t)+B_{\textrm{m}1}^2(t)-A_{\textrm{m}2}^2(t)-B_{\textrm{m}2}^2(t)\right\rangle,
    \label{eq_rad_pres_force}
\end{equation}
where averaging is performed over the period of electromagnetic oscillations. Ignoring the D.C. contribution and linearizing with respect to perturbation terms, the spectrum of the force reads
\begin{align*}
    F_{\textrm{BA}}(\Omega)={}&2\hbar k_0R_\textrm{m}\textit{\textbf{A}}_{\textrm{in}0}^{*T}\mathbb{M}_{A_0}^{*T}\mathbb{M}_b(\Omega)\textit{\textbf{a}}_\textrm{in}(\omega_0+\Omega)\nonumber\\
    &+2\hbar k_0R_\textrm{m}\textit{\textbf{A}}_{\textrm{in}0}^{T}\mathbb{M}_{A_0}^{T}\mathbb{M}_b^*(-\Omega)\textit{\textbf{a}}_\textrm{in}^\dag(\omega_0-\Omega).
\end{align*}
This is the radiation pressure noise, also addressed as back-action noise or stochastic back-action, i.e.~the time-varying radiation pressure that is solely caused by the fluctuations of optical fields. The unsymmetrized spectral density of stationary back-action noise is computed from the equation $2\pi\delta(\Omega-\Omega')S_F(\Omega')=\langle0|F_{\textrm{BA}}(\Omega)F_{\textrm{BA}}^\dag(\Omega')|0\rangle$, yielding
\begin{align}
    S_F(\Omega)={}&\frac{4\hbar k_0}{c}\,\frac{R_\mathrm{m}^2P_\mathrm{in}}{|\mathcal{D}_0\mathcal{D}(\Omega)|^2}
    \left\{\left|\mathfrak{L}(\Omega)\right|^2+T_\mathrm{SR}^2\left|\mathfrak{D}(\Omega)\right|^2\right\},\label{eq_BA_spectr_density}\\
    \mathfrak{L}(\Omega)={}&\alpha_1\left(1+R_\textrm{SR}^2e^{2iK\mathcal{L}}\right)+\alpha_2R_\textrm{SR}e^{2i(k_0+K)\mathcal{L}}\nonumber\\
    &+\alpha_2^*R_\textrm{SR}e^{-2ik_0\mathcal{L}},\nonumber\\
    \mathfrak{D}(\Omega)={}&\beta_1+\beta_2R_\textrm{SR}e^{-2ik_0\mathcal{L}},\nonumber\\
    \alpha_1={}&T_\textrm{m}R_\textrm{BS}T_\textrm{BS}\left(e^{ik\delta l} + e^{-ik\delta l}\right)-R_\textrm{m}(T_\textrm{BS}^2-R_\textrm{BS}^2),\nonumber\\
    \alpha_2={}&T_\textrm{BS}^2e^{ik\delta l} + R_{\textrm{BS}}^2e^{-ik\delta l},\nonumber\\
    \beta_1={}&T_\textrm{m}\left(T_\textrm{BS}^2e^{ik\delta l} - R_{\textrm{BS}}^2e^{-ik\delta l}\right)+2R_\textrm{m}R_\textrm{BS}T_\textrm{BS},\nonumber\\
    \beta_2={}&R_\textrm{BS}T_\textrm{BS}\left(e^{ik\delta l} - e^{-ik\delta l}\right).\nonumber
\end{align}
Here $P_\textrm{in}=\hbar\omega_0|A_{\textrm{L}0}|^2$ is the input laser power, $\mathcal{D}_0=\mathcal{D}|_{k=k_0}$ is the resonant multiplier for mean fields and $\mathcal{D}(\Omega)=\mathcal{D}|_{k=k_0+K}$ is the resonant multiplier for perturbation fields.

\begin{figure}
\begin{center}
\includegraphics[scale=0.45]{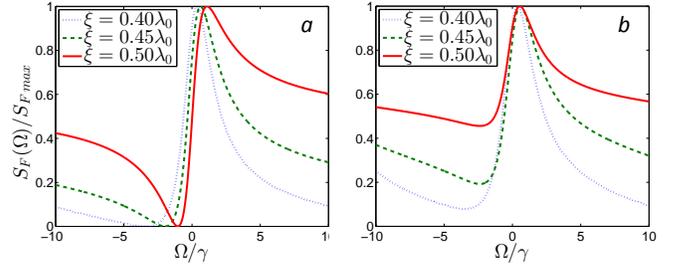}
\caption{Normalized (non-symmetrized) spectral densities of back-action noise for different membrane positions, $\xi=\delta l - \delta l_{\mathrm{DP}}$ and $\delta l_{\mathrm{DP}}$ is defined by Eq.\,(\ref{eq_dark_port_condition}). For better visualization we choose membrane power reflectivity $R_\textrm{m}^2=0.3$, beamsplitter asymmetry $\delta_\textrm{BS}=-0.3$ and detuning $\Delta=0$. \textit{a}: $R_\textrm{SR}^2=1$. \textit{b}: $R_\textrm{SR}^2=0.7$.}
\label{pic_S_F}
\end{center}
\end{figure}

The factors $\mathfrak{L}$ and $\mathfrak{D}$ in Eq.\,(\ref{eq_BA_spectr_density}) describe contributions of vacuum noises from laser ($a_\mathrm{L}$) and detector ports ($a_\mathrm{D}$) respectively. The former one defines the Fano-line profile in the shape of $S_F$ due to interference of input and intracavity laser fields on the membrane, and is identified with the emergence of dissipative coupling in cavity optomechanics \cite{Elste_Girv_Clerk_2009,Xuer_Schab_Hammer_2011,Weiss_Burder_Nunnen_2012}. The latter one, vanishing for the 100\% reflective SRM \cite{Xuer_Schab_Hammer_2011}, describes a Lorentzian profile of the intracavity vacuum field from detector port, and is identified with dispersive coupling in cavity optomechanics. Therefore, spectral density of back-action noise in general case is the mixture of Fano and Lorentz resonances, as mentioned in Sec.\,\ref{sec_canon_anomal_BA}. Plots of normalized $S_F(\Omega)$ are presented in Fig.\,\ref{pic_S_F}\textit{a,b} for $T_\mathrm{SR}=0$ and $T_\mathrm{SR}\neq0$ respectively.

\section{Dynamic back-action}\label{sec_dynamic_BA}

Consider now a movable membrane with position operator $x_\mathrm{m}(t)$ with a corresponding Fourier-transformed operator $x_\mathrm{m}(\Omega)$. According to perturbation theory the fields on the membrane surfaces will have contributions of zeroth and first order in the mechanical displacement. One finds $\textit{\textbf{B}}_{\textrm{m}0}=\mathbb{M}_\textrm{m}\textit{\textbf{A}}_{\textrm{m}0}$ and $\textit{\textbf{b}}_{\textrm{m}}= \mathbb{M}_\textrm{m}\textit{\textbf{a}}_{\textrm{m}}+2ik_0x_\textrm{m}R_\textrm{m}\textit{\textbf{A}}_{\textrm{m}0}$. Thus the perturbation fields now contain both optical noises and the displacement of the membrane. Since the treatment of mean fields remains unchanged, we consider only the perturbation terms. The in-going fields on the beamsplitter are defined by the equation $\textit{\textbf{a}}_\textrm{BS}=\mathbb{P}_\textrm{R}\mathbb{T}_\textrm{R}\textit{\textbf{a}}_\textrm{in}+\mathbb{P}_\textrm{R}\mathbb{R}_\textrm{R}\mathbb{P}_\textrm{R}
\mathbb{M}_\textrm{MS}\textit{\textbf{a}}_{\textrm{BS}}+\mathbb{P}_\textrm{R}\mathbb{R}_\textrm{R}\mathbb{P}_\textrm{R}\mathbb{M}_\textrm{BS}^T\mathbb{P}_L\mathbb{P}_l\,
2ik_0x_\textrm{m}(\Omega)R_\textrm{m}\textit{\textbf{A}}_{\textrm{m}0}$, with solution $\textit{\textbf{a}}_\textrm{BS}=\mathbb{K}_{\textrm{MSR}}\mathbb{P}_\textrm{R}\mathbb{T}_\textrm{R}\textit{\textbf{a}}_{\textrm{in}}
+2ik_0x_\textrm{m}\mathbb{K}_{\textrm{MSR}}\mathbb{P}_\textrm{R}\mathbb{R}_\textrm{R}\mathbb{P}_\textrm{R}\mathbb{M}_\textrm{BS}^T\mathbb{P}_L\mathbb{P}_lR_\textrm{m}
\textit{\textbf{A}}_{\textrm{m}0}$. Thus the incident and reflected fields on the membrane surfaces are
\begin{subequations}
\begin{align}
    \textit{\textbf{a}}_\textrm{m}&=\mathbb{M}_a\textit{\textbf{a}}_{\textrm{in}}+2ik_0x_\textrm{m}\mathbb{M}_{ax}\textit{\textbf{A}}_{\textrm{m}0},
    \label{eq_membr_fields_inc_mov_membr}\\
    \textit{\textbf{b}}_\textrm{m}&=\mathbb{M}_b\textit{\textbf{a}}_{\textrm{in}}+2ik_0x_\textrm{m}(R_\textrm{m}\mathbb{I}+
    \mathbb{M}_\textrm{m}\mathbb{M}_{ax})\textit{\textbf{A}}_{\textrm{m}0}.
    \label{eq_membr_fields_ref_mov_membr}
\end{align}
\end{subequations}
The components of matrix
\begin{equation*}
    \mathbb{M}_{ax}=\mathbb{P}_l\mathbb{P}_L\mathbb{M}_\textrm{BS}\mathbb{K}_{\textrm{MSR}}\mathbb{P}_\textrm{R}\mathbb{R}_\textrm{R}\mathbb{P}_\textrm{R}
    \mathbb{M}_\textrm{BS}^T\mathbb{P}_L\mathbb{P}_lR_\textrm{m}.
\end{equation*}
are:
\begin{align*}
    \mathbb{M}_{ax}^{(1,1)}&=\mathcal{D}^{-1}R_\textrm{m}R_\textrm{BS}^2R_\textrm{SR}e^{2ik(\mathcal{L}-\delta l/2)},\\
    \mathbb{M}_{ax}^{(2,2)}&=\mathcal{D}^{-1}R_\textrm{m}T_\textrm{BS}^2R_\textrm{SR}e^{2ik(\mathcal{L}+\delta l/2)},\\
    \mathbb{M}_{ax}^{(1,2)}&=\mathbb{M}_{ax}^{(2,1)}=-\mathcal{D}^{-1}R_\textrm{m}R_\textrm{BS}T_\textrm{BS}R_\textrm{SR}e^{2ik\mathcal{L}}.
\end{align*}

Substituting mean fields from Eqs.\,(\ref{eq_membr_fields_inc_immov_membr},\,\ref{eq_membr_fields_ref_immov_membr}) and perturbations fields (\ref{eq_membr_fields_inc_mov_membr},\,\ref{eq_membr_fields_ref_mov_membr}) into Eq.\,(\ref{eq_rad_pres_force}), ignoring the D.C. part and linearizing with respect to perturbation terms, one ends up with $F(\Omega)=F_\mathrm{BA}(\Omega)+F_x(\Omega)$. Here $F_\mathrm{BA}$ is the radiation pressure noise considered in Appendix \ref{sec_stochastic_BA}, and $F_x(\Omega)=-\mathcal{K}(\Omega)x_\mathrm{m}(\Omega)$ is the ponderomotive force, i.e. dynamical part of the radiation pressure force caused by the motion of the membrane. The coefficient $\mathcal{K}(\Omega)$ modifies the dynamics of the membrane, and therefore represents the dynamic back-action,
\begin{align*}
    \mathcal{K}(\Omega)&=\frac{2ik_0}{c}\,R_\textrm{m}P_\textrm{in}\left[\mathbb{K}_{(1,1)}(\Omega)-\mathbb{K}_{(1,1)}^*(-\Omega)\right],\\
    \mathbb{K}(\Omega)&=\mathbb{M}_{B_0}^{*T}\left[\sigma_3-2\mathbb{M}_{ax}(\Omega)\right]\mathbb{M}_{A_0},\nonumber
\end{align*}
with $\sigma_3=\mathrm{diag}(1,-1)$. If one denotes $K(\Omega)\equiv\Re[\mathcal{K}(\Omega)]$ and $\Gamma(\Omega)\equiv-\frac{1}{2}\Im[\mathcal{K}(\Omega)]/\Omega$, then the corresponding shifts of the square of intrinsic mechanical frequency and damping rate are equal to $K/m$ and $\Gamma/m$ \footnote{Assuming that the equation of motion of the membrane is of the following form: $\ddot{x}+2\gamma_\mathrm{m}\dot{x}+\omega_\mathrm{m}^2x=F(t)/m$.}. After applying simplifying conditions described in Sec.\,\ref{sec_canon_anomal_BA}, one ends up with Eq.\,(\ref{eq_dynamic_BA_gen_approx}) for $\mathcal{K}$ presented there.

\addcontentsline{toc}{section}{References}

%

\end{document}